\documentclass[12pt,preprint,a4paper]{aastex}
\usepackage{times}

\usepackage{amsmath}                % American Mathematical Society package
\usepackage{amsfonts}               % American Mathematical Society fonts
\usepackage{amssymb}                % American Mathematical Society symbol
\usepackage{natbib}
\usepackage{graphicx}
\usepackage[tight]{subfigure}
\usepackage{times}
\usepackage{nicefrac}

\def \cm{~\rm{cm}}
\def \s{~\rm{s}}
\def \km{~\rm{km}}

\def \K{~\rm{K}}
\def \g{~\rm{g}}

\def \erg{~\rm{erg}}

\def \yr{~\rm{yr}}

\title{LIMITS ON CORE DRIVEN ILOT OUTBURSTS OF ASYMPTOTIC GIANT BRANCH STARS}

\author{Liron Mcley\altaffilmark{1} and Noam Soker\altaffilmark{1}}

 \altaffiltext{1}{Department of Physics, Technion -- Israel Institute of
Technology, Haifa 32000 Israel; lironmc@tx.technion.ac.il, soker@physics.technion.ac.il.}

\begin{document}

\begin{abstract}
We find that single-star mechanisms for Intermediate Luminosity Optical Transients (ILOTs; Red Transients; Red Novae) which are powered by energy release in the core
of asymptotic giant branch (AGB) stars are likely to eject the entire envelope, and hence cannot explain ILOTs in AGB and similar stars.
There are single-star and binary models for the powering of ILOTs, which are eruptive stars with peak luminosities between those of novae and supernovae.
In single-star models the ejection of gas at velocities of $\sim 500-1000 \km \s^{-1}$ and a possible bright ionizing flash,
require a shock to propagate from the core outward.
Using a self similar solution to follow the propagation of the shock through the envelope of two evolved stellar models, a 
$6M_\odot$ AGB star and an $11 M_\odot$ yellow supergiant (YSG) star, we find that the shock that is required to explain the observed mass loss
also ejects most of the envelope.
We also show that for the event to have a strong ionizing flash the required energy expels most of the envelope.
The removal of most of the envelope is in contradiction with observations.
We conclude that single-star models for ILOTs of evolved giant stars encounter severe difficulties.
\end{abstract}

% ==========================================================
\section{INTRODUCTION}
\label{sec:intro}
% ==========================================================

The eruptive stars with peak luminosity between those of novae and supernovae
(e.g. \citealt{Mould1990, Rau2007, Prieto2009, Ofek2008, Botticella2009, Smithetal2009, Berger2009, KulkarniKasliwal2009, Mason2010,
Pastorello2010, Kasliwaletal2011, Tylendaetal2013})
form a heterogenous group \citep{Kasliwal2011} with many enigmatic objects, and with no consensus on how to name them.
We will refer to objects in this group as Intermediate-Luminosity Optical Transients (ILOTs; \citealt{Berger2009}),
but note that Red Novae, Optical Transients, and Red Transients are also in common use, for at least some of these objects.
Not only is the name for these objects not in consensus,
but also the powering processes of many of them and whether they are due to binary interaction or
are formed through single star evolution are debated.
 Since this is a heterogenous group, it may consist of several subtypes differing by their origin and cause.
The focus of this study is ILOTs whose pre-outburst objects are asymptotic giant branch (AGB) or extreme-AGB (EAGB) stars,
e.g., NGC~300~OT2008-1 (NGC~300OT; \citealt{Monard2008, Bond2009, Berger2009}) and SN~2008S \citep{ArbourBoles2008}.

There are single star models (e.g., \citealt{Thompsonetal2009, Kochanek2011}) and binary stellar models
\citep{Kashi2010, KashiSoker2010b, SokerKashi2011, SokerKashi2012, SokerKashi2013} for ILOT events harboring AGB stars.
The single star mechanisms that were listed by \cite{Thompsonetal2009} for EAGB-ILOTs are the formation of massive ONeMg WD,
electron capture supernovae (ecSNe; also \citealt{Botticella2009} for SN 2008S), iron core collapse supernovae (CCSNe), and
outbursts of massive $\sim 10-15 M_\odot$ stars.
\cite{Bond2009} derived the total radiated energy in the NGC~300OT outburst to be $\sim 10^{47} \erg$,
and concluded that the NGC 300OT was an eruption of a $\sim 10-15 M_\odot$ star that cleared the surrounding dust and initiated a bipolar wind.
They mention an unexplained failed SN, a binary merger, or a photospheric eruption, as possible mechanism for the NGC~300OT outburst.
\cite{Berger2009} found velocities in the range of $\sim 200 -1000 \km \s^{-1}$ in the outburst of NGC~300OT, and concluded based on these,
the luminosity, and an overall similarity to the ILOTs SN~2008S \citep{Prieto2009} and M85~OT2006-1 that the outburst did not result in a
complete disruption of the progenitor.

\cite{Thompsonetal2009} mentioned that the formation of a massive ONeMg WD in the single-star scenario for EAGB-ILOTs might form a bipolar PNe.
The connection between these ILOTs and bipolar PNs was mentioned also in the binary model for EAGB-ILOTs \citep{SokerKashi2012, AkashiSoker2013}.
\cite{Prieto2009} already made a connection between NGC~300OT and pre-PNe, and raised the possibility that the progenitor of NGC~300OT was
of mass $< 8 ~\rm{M_\odot}$.
Recently more PNe were suggested to have part of their nebula ejected in an ILOT event, e.g., KjPn~8 \citep{BoumisMeaburn2013}.
In the binary model a companion star, mostly a main sequence (MS) star, accretes part of the mass ejected by the evolved star.
The gravitational energy released is channelled directly to radiation and to kinetic energy of the ejected mass.
The interaction of the gas ejected in the event with previously ejected gas further increases the radiated energy.
A large fraction of the ejected gas can reside in jets launched by the accreting companion star.
The jets lead to the formation of a bipolar nebula, such as the Homunculus--the bipolar nebula of Eta Carinae \citep{KashiSoker2010a}--that was
formed in the nineteenth century Great Eruption of Eta Carinae.

\cite{Kochanek2011} performed a thorough analysis of the dust destruction by the shock breakout luminosity of SN~2008S and NGC~300OT,
and the dust reformation.
  (The `shock breakout' moment refers to the time when photons behind the shock expand freely to the observer. This occurs when the shock front is
near the photosphere.)   
\cite{Kochanek2011} concluded that the progenitors were red supergiants, and found that the required shock breakout luminosities were of order $10^{10} L_\odot$,
and hence the outbursts in both systems were explosive in nature.
However, he could not tell whether the progenitors survived the outburst.
Based on \cite{Berger2009},  \cite{Smithetal2011} adopted an expansion speed of $v_e \simeq 560 \km \s^{-1}$ for NGC~300OT.
\cite{Smithetal2009} noted expansion speeds of about $v_e \simeq 600- 1000 \km \s^{-1}$ in SN~2008S, and estimated
that it ejected $M_e \simeq 0.05 - 0.2 M_\odot$ in the outburst, and radiated $\sim 6 \times 10^{47} \erg$.
\cite{Kochanek2011} estimated the ejected mass and radiated energy to be $M_e > 0.05 M_\odot$ and $M_e >0.25 M_\odot$,
and $\sim 3 \times 10^{47} \erg$ and $\sim 8 \times 10^{47} \erg$, for SN~2008S and NGC~300OT, respectively.

In this study we examine the implications of such outbursts which are powered from within the stellar core.
We aim at answering the following questions.
(1) What are the typical energies required to account for the observed outbursts? This question is motivated by
the binary-model for these outbursts where the energy comes from accretion onto a companion. In the binary model a large fraction of the
liberated energy goes directly to radiation or to kinetic energy which is later channelled to radiation by wind collisions.
We will find that the efficiency of the single-star model is very low.
(2) Can such outbursts leave a large fraction of the envelope bound?
Namely, does the star survive the outburst? This question is also motivated by the binary-model,
where most of the envelope of the primary star stays bound.

In section \ref{sec:ejection} we derive the total energy that is deposited in the envelope from requirements on the ejected gas, and find that most
of the envelope must be ejected to account for the velocities of the ejected mass.
Our findings are similar to those of \cite{Dessartetal2010} who performed 1D simulations of the same process for $10-25 M_\odot$ stellar models.
In section \ref{sec:shockb} we derive the energy deposited to the envelope from the required luminosity during the shock breakout, and
our short summary is in section \ref{sec:summary}.

% ==========================================================
\section{EJECTING HIGH VELOCITY GAS}
\label{sec:ejection}
% ==========================================================

% =======================
\subsection{The self similar calculation and models}
\label{subsec:self}
% =======================

\begin{figure}[ht]
  \centering
       \includegraphics[width=0.45\textwidth]{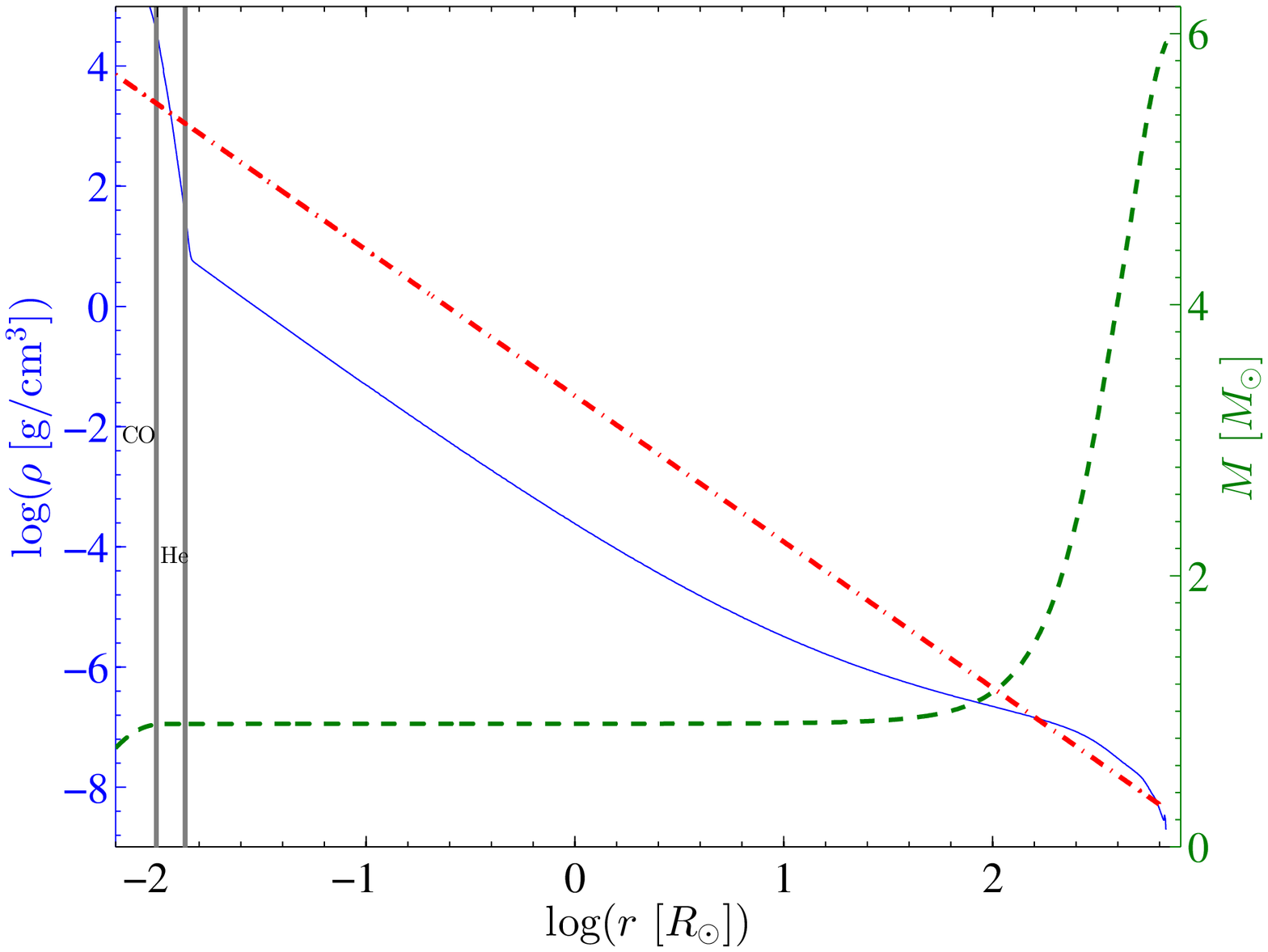}
       \hskip 0.6cm
       \includegraphics[width=0.45\textwidth]{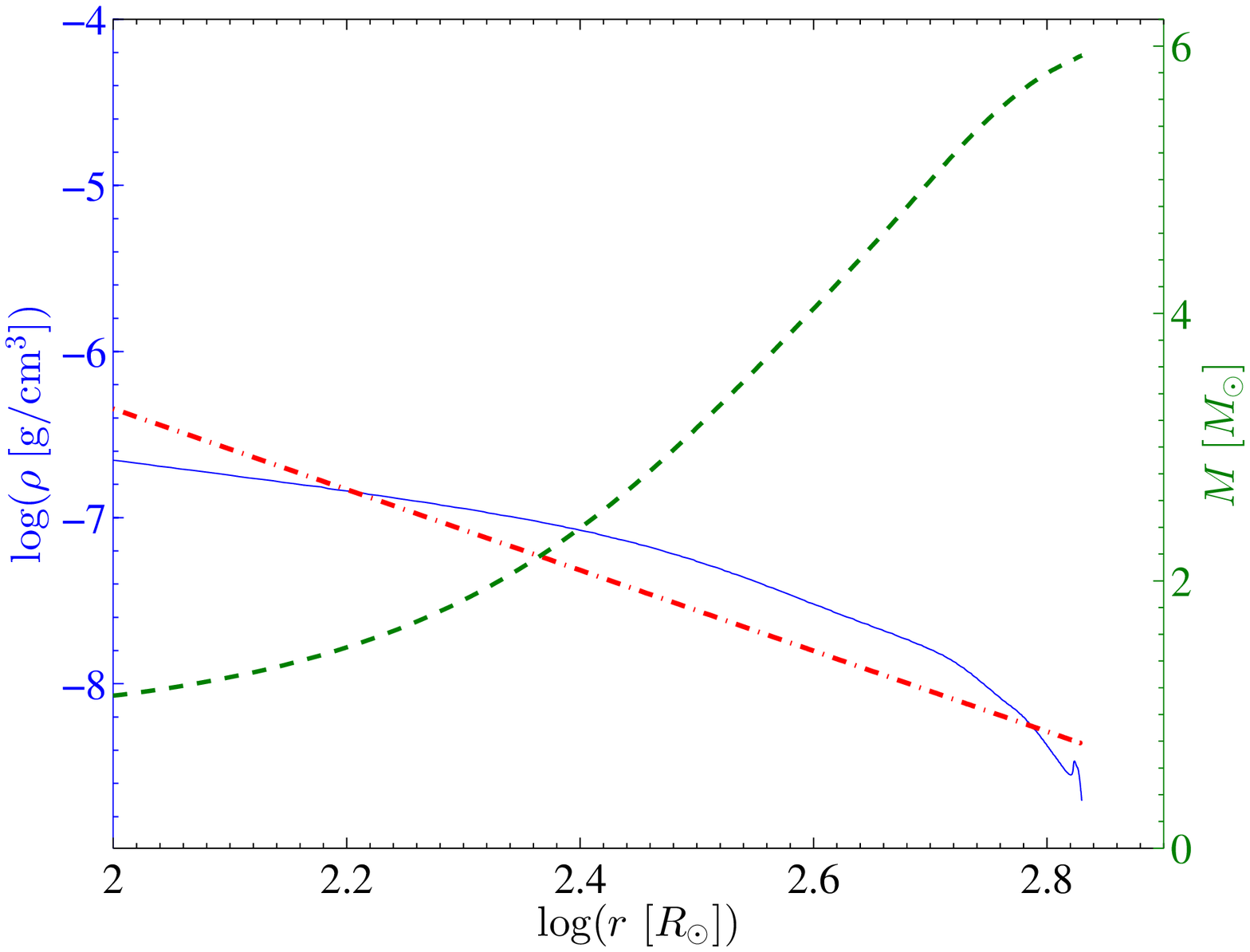}
       \caption{The $6 M_{\odot}$ AGB model calculated with MESA. The stellar radius and luminosity are $R_{\rm *} = 675 R_\odot$
       and $L_{\rm *} = 4.2 \times 10^4 L_\odot $, respectively.
       The blue solid line and the green dashed line represent the density and mass profiles of the model, respectively.
The right panel enlarges the outer envelope.
 The red dash-dot line represents our power law fitting to the outer region of the envelope as given by equation (\ref{eq:density1}).
 We none-the-less, use this
 fitting for the entire envelope at $r \ga 0.038 R_\odot$ in our self-similar solution.    }
       \label{fig:AGBm}
\end{figure}

\begin{figure}[ht]
  \centering
       \includegraphics[width=0.7\textwidth] {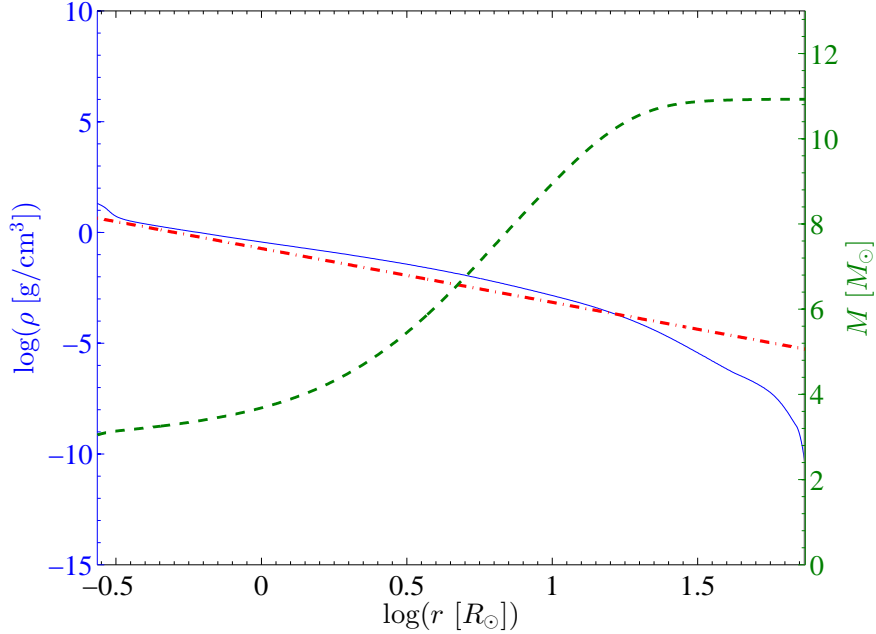}
       \caption{Like figure \ref{fig:AGBm} but for the $11 M_{\odot}$ yellow supergiant (YSG) model. The stellar radius is $R_{\rm *} = 74 R_\odot$ and its luminosity is $L_{\rm *} = 2.6\times 10^4 L_\odot$.  }
       \label{fig:YSGm}
\end{figure}

We use two models calculated with the stellar evolution code MESA \citep{Paxton2011}.
Both calculations were done for non-rotating stars with solar metallicity ($Z=0.02$) at
zero-age main sequence (ZAMS).
 The masses of $ 6M_\odot$ and $11 M_\odot$ were chosen to sample the two regions below and above the critical main sequence mass 
for supernova explosions, $\sim 8.5 M_\odot$. 
Figs. \ref{fig:AGBm} and \ref{fig:YSGm} display the density and mass profiles of the two models.
The AGB stellar model (Fig. \ref{fig:AGBm}) has a ZAMS mass of $6 M_\odot$. We take it at its AGB phase, at an age of $7.1 \times 10^{7} \yr$,
when its radius is $R_{\rm *} = 675 R_\odot$, its luminosity is $L_{\rm *} = 4.2 \times 10^4 L_\odot$, its effective temperature is $T_{\rm *} = 3.2\times 10^{3} \K$,
it has developed a $0.90 M_\odot$ CO core, and a $ 0.0048 M_\odot$ He mantle (outer core), and maintained a total mass of $5.98 M_\odot$.
The yellow supergiant (YSG) stellar model (Fig. \ref{fig:YSGm}) has a ZAMS mass of $11 M_\odot$. We take it during
its helium shell burning phase, at an age of $ 2 \times 10^{7} \yr$, when its radius is $R_{\rm *} = 74 R_\odot$, its luminosity is $L_{\rm *} = 2.6\times 10^4 L_\odot$,
its effective temperature is $T_{\rm *} = 8.5\times 10^3 \K$, it has developed a $ 1.5 M_\odot$ CO core, the outer boundary of the helium shell
is at $M=3 M_\odot$, and the total stellar mass is $10.9 M_\odot$.
% ====================
% ====================
% ====================
% ====================

To facilitate an analytical self-similar solution we make a number of assumptions as follows.
(1) The flow is spherically symmetric.
(2) The envelope density profile can be approximated as a power law given by
\begin{equation}
\rho(r) =Br^{-\omega} =
\begin{cases}
    0.0325 \left(\frac{r}{R_\odot} \right)^{-17/7} \g \cm^{-3}, \quad M_{\rm ssA}=5.04 M_\odot, \quad & \text{AGB}\\
    0.1881 \left(\frac{r}{R_\odot} \right)^{-17/7} \g \cm^{-3}, \quad M_{\rm
     ssY}=8.2M_\odot, \quad & \text{YSG}   ,
  \end{cases}
\label{eq:density1}
\end{equation}
where $M_{\rm ss}$ is the total mass used in the self similar solution. This mass is somewhat larger than the envelope
mass because the density profile starts at $r=0$. The envelope masses are $M_{\rm env}(AGB)= 5.02 M_\odot$ and
$M_{\rm env}(YSG)= 7.9 M_\odot$.
This power-law density profile is presented with the red dash-dot line in Fig. \ref{fig:AGBm} and \ref{fig:YSGm}.
(3) The explosion energy $E$ originated from the core region and within a time much shorter than the
dynamical time in the envelope. This assumption allows us to treat the explosion as an instantaneous release of energy
from the center of symmetry.
(4) We assumed that radiation pressure dominates the thermal pressure so that the gas can be treated as a $\gamma=4/3$ fluid.
Although this assumption is not accurate for slow shocks, we will nevertheless present results for relatively slow shocks as well,
where the gas pressure is higher than the radiation pressure, in order to estimate the outcome.
In section \ref{subsec:solution} below we argue that higher values of $\gamma$ will only strengthen our conclusions.
(5) As the pre-shock pressure, i.e., the pressure in the envelope, is much lower than the post-shock pressure,
we neglect the pre-shock pressure altogether.
These assumptions imply that the solution asymptotically approaches the self similar solution for a shock wave.

We note that the power-law density profile given in equation (\ref{eq:density1}) does not fit the inner part and the very outer part
of the AGB envelope (Fig. \ref{fig:AGBm}), and the outer part of the YSG envelope (Fig. \ref{fig:YSGm}).
We discuss the implications of the mismatch in the very outer part of the envelope in section \ref{subsec:solution}.

\begin{figure}[ht]
  \centering
        \includegraphics[width=0.65\textwidth]{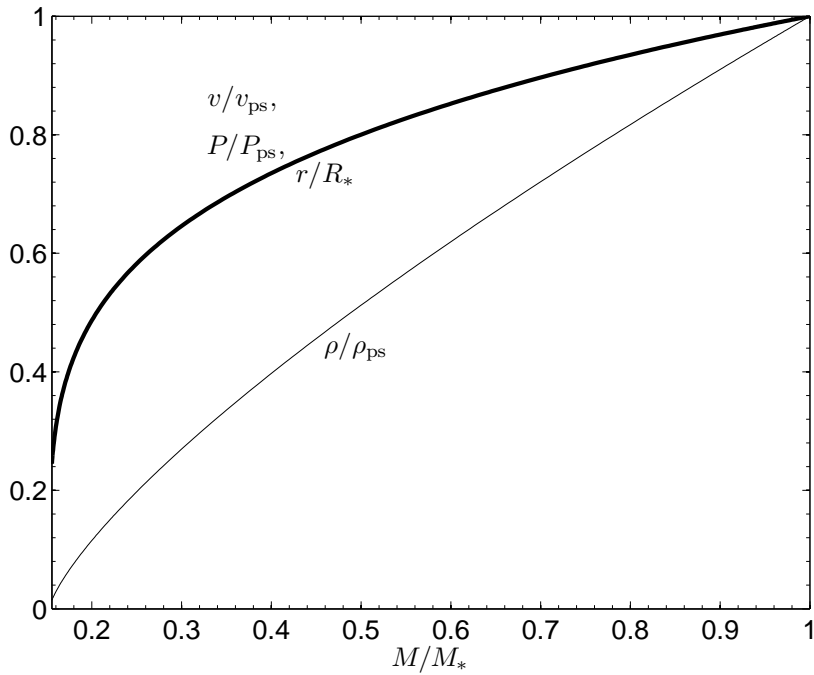}
       \caption{The self similar solution for $\gamma =4/3$ and $\omega=17/7$ .
       The thick upper line represents the radius relative to the stellar radius, and the velocity and pressure relative
       to their immediate post-shock values; all these coincide. The lower line represents the density relative to its immediate post-shock value. }
\label{fig:SelfSim}
\end{figure}

We turn now to present the self similar solution to the shock.
We follow the self similar solution as described by \cite{Chevalier1976},
for a shock wave travelling through an envelope with a density profile given by equation (\ref{eq:density1});
see corrections in  \cite{ChevalierSoker1989}.
The equations describing the post shock variables when the shock reaches $R_{\rm *}$ are,
% Fit the Chevalier equations with what I've done!!! Check that it's okay!!
\begin{equation}
   v=\frac{2}{3}\frac{r}{t_{\rm end}}\;\;,\;\;e_{\rm k}=\frac{1}{2}v^2 ,
 \label{eq:SelfSimVar1}
\end{equation}
\begin{equation}
   \rho =7BA^{-4/3}rt^{-8/3}_{\rm end} ,
 \label{eq:SelfSimVar2}
\end{equation}
\begin{equation}
   P=\frac{14}{27}BA^{-4/3}r^3 t^{-14/3}_{end}    \;\;,\;\;e_{\rm int}=\frac{1}{\gamma -1}\frac{P}{\rho} ,
\label{eq:SelfSimVar3}
\end{equation}
\begin{equation}
%%%        t_{end}=\left[\left(\frac{A\alpha}{E_0}\right)^{{(5-\omega)}}R_{star}\right]^{{(5-\omega)}{2}}
t_{\rm end}= R_{\rm *} ^{9/7} A^{-1/2},
\label{eq:SelfSimVar4}
\end{equation}
where $e_{\rm k}$ and $e_{\rm int}$ are the specific kinetic and internal energies, respectively, $t_{\rm end}$ is the time when
the shock reaches the radius of the star $R_{\rm *}$, and $A=27E/(56\pi B)$ is determined by requiring energy conservation, $E$ being the total energy deposited in the envelope.
For the solutions presented here the adiabatic index is $\gamma =4/3$ and for $\omega=17/7$ used here $e_{\rm k} = e_{\rm int}$
in the post-shock region.
The self similar solution of the post-shock region when the shock just reaches the stellar radius is presented
in Fig. \ref{fig:SelfSim}.
% ====================
% ====================

% =======================
\subsection{Results}
\label{subsec:solution}
% =======================

We aim at examining how much energy is left in the envelope after the shock had passed, and in particular
what fraction of the envelope has a positive energy, and hence will be ejected.
The constraints on the solution from observations, as discussed in section \ref{sec:intro},
are that the shock should accelerate a mass of $M_e \sim 0.05-0.5 M_\odot$ to velocities of $v_e \simeq 500-1000 \km \s^{-1}$.
We note that there is a steep density gradient at the edge of the envelope, with a slope much larger than $\omega=17/7$.
In such a steep envelope a shock can accelerate gas to velocities of $\sim 3$ times the initial shock velocity (e.g., \citealt{TsebrenkoSoker2013}).
Such an acceleration in a steep density gradient has been suggested by \cite{Humphreysetal2012} to explain SN~2011ht as a SN impostor. 
For that reason we consider a solution with shock velocities down to $v_s = 200 \km \s^{-1}$.
In that case the post-shock gas pressure dominates the radiation pressure, and taking the adiabatic index to be $\gamma =4/3$ gives only
a crude solution. However, this is adequate to our purposes and the other approximations, e.g., the stellar structure.
Moreover, since the adiabatic index corresponds to the number of internal degrees of freedom, a larger adiabatic index,
as would be the case when thermal pressure dominates, implies that more energy is available for expelling the envelope of the star.

The results of \cite{TsebrenkoSoker2013} show that the mass accelerated to high velocities in a steep density gradient is a small
fraction of the gas mass travelling at lower velocities. For example, to eject $\sim 0.01 M_\odot$ at $\sim 500 \km \s^{-1}$ by post-shock gas
at velocity of $\sim 250 \km \s^{-1}$, we need $\ga 1 M_\odot$ of gas moving at $\sim 250 \km \s^{-1}$.

The initial density profile of the self similar solution given in equation (\ref{eq:density1}) does not include
the core mass. The core mass $M_{\rm core}$ must be included in calculating the gravitational energy of the envelope.
For that, in calculating the specific gravitational energy of a mass element at radius $r$, $e_{G}(r)$, we add the core mass $M_{\rm core}$ to the
self-similar mass inner to $r$, $M_{\rm ss}(r)$.
For the same reason, our approach to the self-similar solution cannot deal with the mass that started within the core radius.
We therefore present results just for the gas that started in the envelope, and the graphs below do not start at $M=0$.

\begin{figure}[ht]
   \centering
      \includegraphics[width=0.49\textwidth]{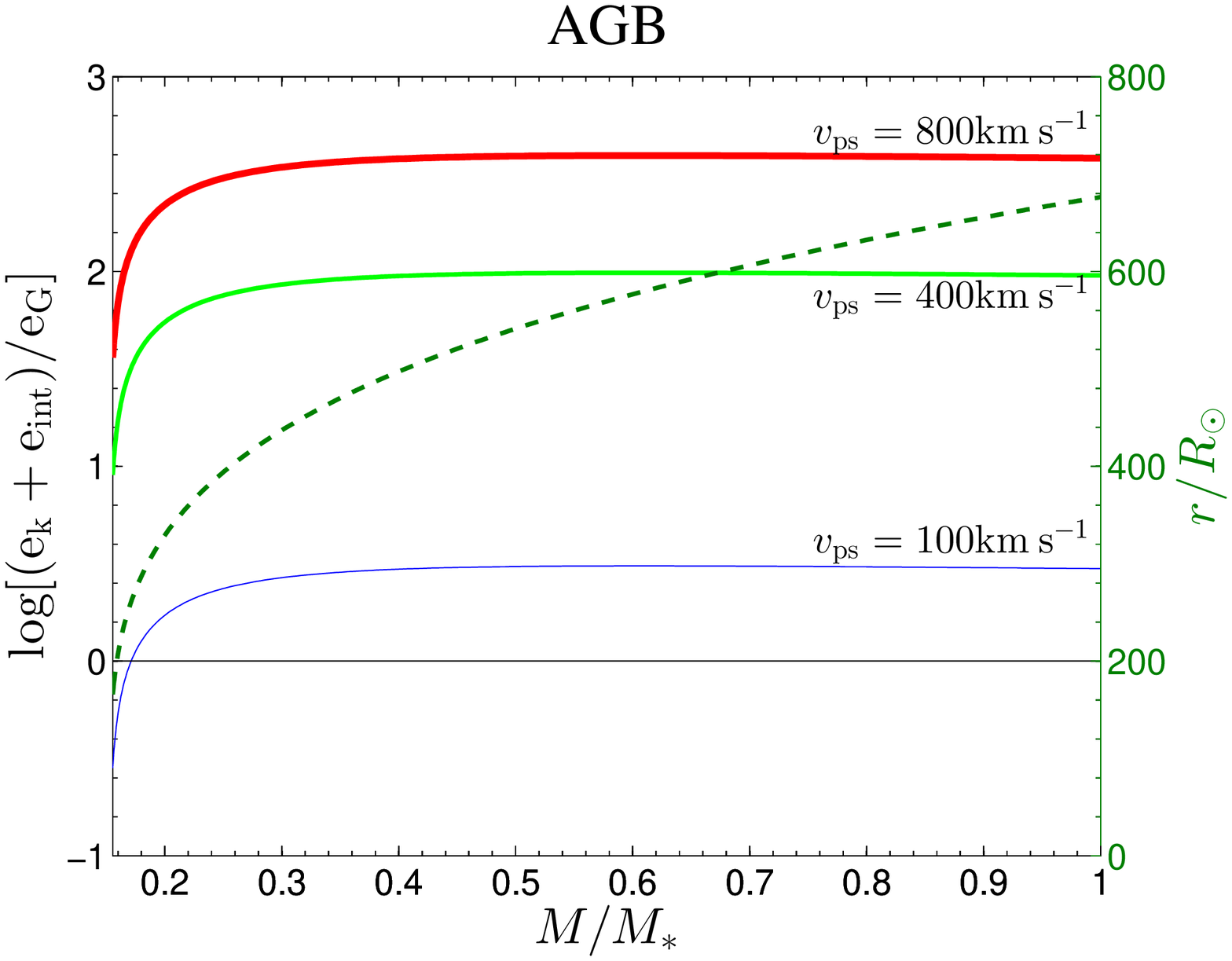}
      \hskip 0.2cm
       \includegraphics[width=0.49\textwidth]{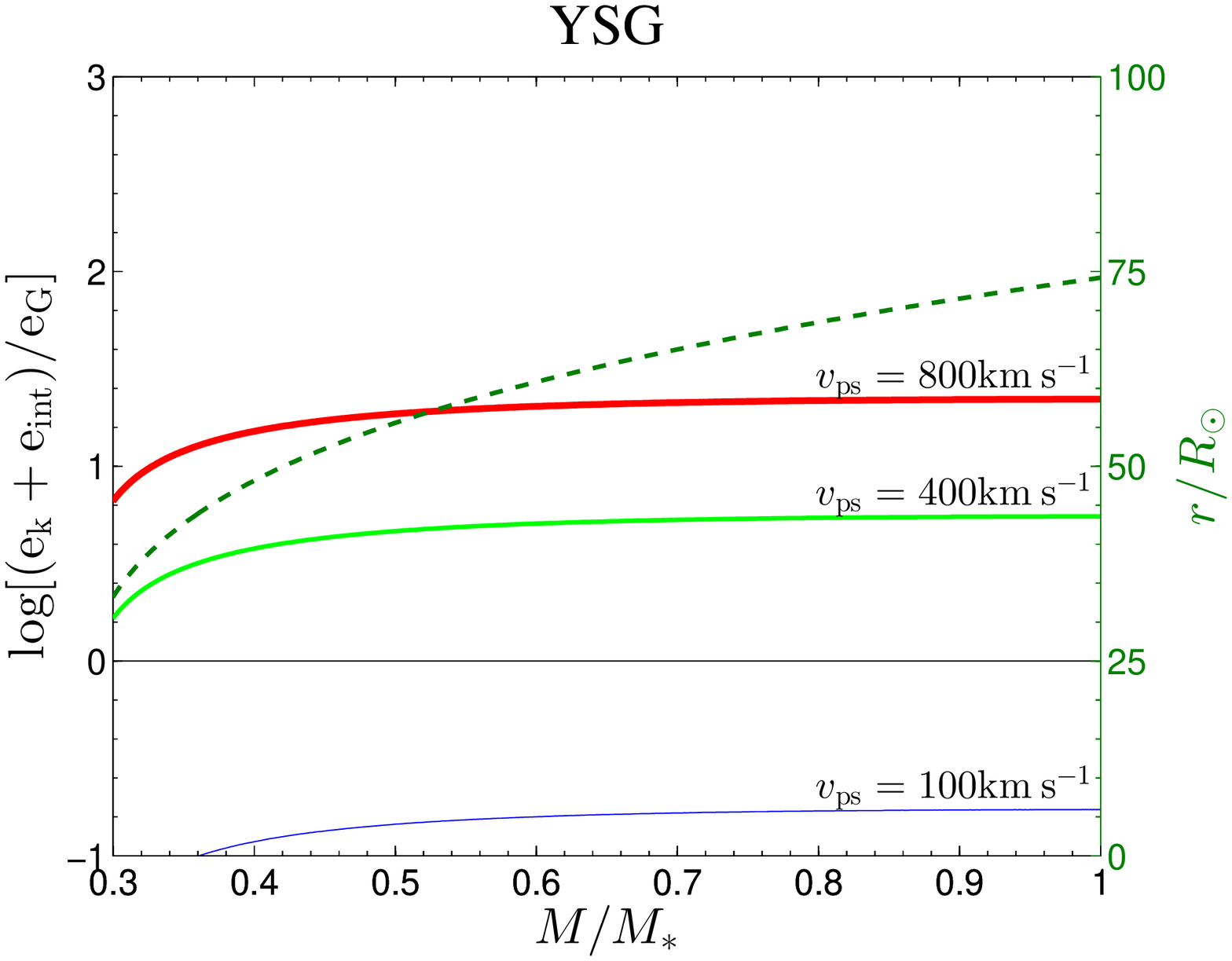}
      \caption{The specific energy of the post-shock stellar envelope gas relative to its binding energy, $\xi(m)=(e_{k} + e_{\rm int})/e_{G}$, as function of mass coordinate
      in the envelope, and for three values of the immediate post-shock velocity at the stellar surface, $v_{\rm ps} = 100$, $400$ and $800 \km \s^{-1}$.
      $\xi(m)$ is given in logarithmic scale as marked on the left axis.
      Results for the AGB and YSG models are presented on the left and right, respectively.
      The horizontal line in each panel marks $\xi =1$. Gas elements above this line are unbound.
     The dashed line is the radius as function of mass in the envelope, with its scale marked on the right axis.}
      \label{fig:EkinEint-Eg}
\end{figure}

In figure \ref{fig:EkinEint-Eg} we present results characterized by the immediate post-shock velocity of the gas at the stellar surface $v_{\rm ps}$.
The energy deposited in the center in our calculation is
\begin{equation}
E_{\rm AGB} =1.68 \times 10^{49} \left( \frac{v_{\rm ps}}{500 \km \s^{-1}} \right)^2 \erg
\label{eq:energyAGB}
\end{equation}
for the AGB model and
\begin{equation}
E_{\rm YSG} =2.73 \times 10^{49} \left( \frac{v_{\rm ps}}{500 \km \s^{-1}} \right)^2 \erg
\label{eq:energyYSG}
\end{equation}
for the YSG model. These values where calculated using equations \ref{eq:SelfSimVar1}-\ref{eq:SelfSimVar4} so that matter at $R_{\rm *}$ would acquire
$v_{\rm ps}$ when the shock reaches the stellar radius.
For each of the three values of $v_{\rm ps}$ we show the ratio of the specific gas energy to its specific binding energy, $\xi (m) = (e_{\rm k} + e_{\rm int})/e_{\rm G}$,
as function of mass coordinate in the envelope.
The specific binding energy is given by $e_{\rm G} = G[M_{\rm core} + M_{\rm env} (r)]/r$, and $e_{k}$ and $e_{\rm int}$ are the specific kinetic and internal energies.
When $\xi > 1$ the gas is basically unbound.
%%% In section \ref{sec:summary} we discuss the implication of the almost horizontal shape of the $\xi(m)$ profile.
 The almost horizontal shape of the $\xi(m)$ profile implies that almost the entire envelope is ejected. Our results are similar to those of
\cite{Dessartetal2010} who performed 1D simulations of the same process for $10-25 M_\odot$ stellar models. \cite{Dessartetal2010} were interested mainly in pre-SN mass ejection,
and did not refer much to surviving stars.
While the explosive mass ejection by energy deposition near the core might work in major envelope removal, our results show that it cannot account for stars with
repeated ILOT events, e.g., $\eta$ Car, as we discuss in section \ref{sec:summary}.

% ==========================================================
\section{SHOCK BREAKOUT}
\label{sec:shockb}
% ==========================================================

In this section we consider the difficulty of a single star model powered by the stellar core from another direction.
In his shock breakout model for NGC~300-OT and SN~2008S \cite{Kochanek2011} requires the transient to be an explosive event,
with a peak luminosity of $L_{\rm peak} \sim 3 \times 10^{10} L_\odot$.
The stellar radius in the model considered by \cite{Kochanek2011} is $R_\ast \simeq 5 \times 10^{13} \cm$, and the
duration of the breakout peak is of the order of the light crossing time $R_\ast / c \la 10^4 \s$.

We use the parameters from \cite{Kochanek2011} in the breakout shock solution of \cite{Katzetal2012}, and derive the required energy for the
explosion.
The peak luminosity of the shock breakout is given by equation (9) of \cite{Katzetal2012}
\begin{equation}
L_{\rm peak}  \simeq 0.33 4 \pi R_\ast^2 \rho _0 v_0^3, \label{eq:L1}
\end{equation}
while the typical duration is (their eq. 5)
\begin{equation}
t_0 \simeq \frac {c}{\kappa \rho_0 v_0^2} .
\label{eq:t0}
\end{equation}
Here $v_0$ is the shock velocity, $\rho_0$ is the pre-shock density, and $\kappa$ is the opacity.
Last two equations give
\begin{equation}
L_{\rm peak}  \simeq  4 R_\ast^2  v_0 c (\kappa t_0 )^{-1} .
\label{eq:L2}
\end{equation}
Their  derivation gives for a $5 M_\odot$ envelope (their eq. 14)
\begin{equation}
v_0 \simeq 2000 \left( \frac{E_{\rm in}}{10^{50} \erg} \right) ^{0.5} \km \s^{-1},
\label{eq:v0}
\end{equation}
where $E_{\rm in}$ is the total energy injected into the envelope in the explosion.
From equations (\ref{eq:L2}) and (\ref{eq:v0}) and for Thomson scattering opacity we derive
\begin{equation}
L_{\rm peak}  \simeq  4 \times 10^{10}
\left( \frac{E_{\rm in}}{10^{50} \erg} \right) ^{0.5}
\left( \frac{t_0}{1000 \s} \right)^{-1}
\left( \frac{R}{5\times 10^{13} \cm} \right)^2
L_\odot .
\label{eq:L3}
\end{equation}

We conclude that in order to have an outburst with the luminosity required by the model of \cite{Kochanek2011}, the explosion energy should be
$E_{\rm in} \sim 10^{50} \erg$.
This energy is much larger than the binding energy of the envelope of EAGB stars, resulting in a massive mass ejection from the envelope.
Basically, the entire envelope becomes unbound by the outburst.

% ==========================================================
\section{IMPLICATIONS AND SUMMARY}
\label{sec:summary}
% ==========================================================

We used a self-similar solution to study energy deposition within a short time scale near evolved stellar cores.
This process is at the heart of some single-star models for the outburst of intermediate luminosity optical transients (ILOTs).
The two stellar models and our approximated power-law density profiles are shown in Fig. \ref{fig:AGBm} and \ref{fig:YSGm}.
A shock runs through the stellar envelope, as depicted in Fig. \ref{fig:SelfSim}. In Fig. \ref{fig:EkinEint-Eg} we present the ratio of gas energy to its binding energy as a
function of mass coordinate in the envelope $\xi(m)$, when the shock front reaches the stellar surface.

There is one clear implication from the behavior of $\xi(m)$: To eject a small amount of mass from the surface at more than the escape speed,
implies that almost the entire envelope is lost.
Our results are similar to those of \cite{Dessartetal2010} who were aiming indeed at explaining massive envelope removal.
Namely, the star is seriously disrupted, it cannot return to its previous structure, and also cannot experience repeated ILOT events.

Very small amounts of gas, below those required by the observations described in section \ref{sec:intro}, can be ejected because of the steep
density profile on the stellar surface. Typically, at velocities of three times the shock velocity (e.g., \citealt{TsebrenkoSoker2013}).
The results of \cite{TsebrenkoSoker2013} show that the mass accelerated to high velocities in a steep density gradient is a very small
fraction of the gas mass at lower velocities. For example, to eject $\sim 0.1 M_\odot$ at $\sim 500 \km \s^{-1}$ by post-shock gas
at velocity of $\sim 250 \km \s^{-1}$, we need $\ga 10 M_\odot$ of gas moving at $\sim 250 \km \s^{-1}$. Again, this implies that most of the
envelope is ejected.

In section \ref{sec:shockb} we derived the energy that is required to be deposited in the core to obtain the luminosity in the
shock breakout model of \cite{Kochanek2011} for the ILOTs NGC~300-OT and SN~2008S.
We find that the required energy is $\sim 10^{50} \erg$, as given by equation (\ref{eq:L3}).
Again, the entire envelope is lost by such an energy deposition.
This strengthens the results we have found in section \ref{subsec:solution}.
Namely, to account for energetic ILOTs by energy deposited near the core, the entire envelope is lost. This is in contradiction
to observations presented in section \ref{sec:intro}, and to some ILOTs that seem to return to their previous stage, e.g.,
$\eta$ Carinae after the Great Eruption \citep{DavidsonHumphreys1997} and the first two outbursts of SN~2009ip \citep{Mauerhanetal2013}.
However, if it turns out that in an ILOT most of the envelope is ejected, then energy deposition from the core is required,
and indeed can account for such a process \citep{Dessartetal2010}.

The problems we find with models based on a shock running through the envelope, as required for a shock breakout UV flash, do not apply to the case
were the core excites waves propagating throughout the envelope, as in the wave-driven mass loss in the SN progenitor
mechanism proposed by \cite{QuataertShiode2012} and \cite{ShiodeQuataert2014}.
The wave-driven mass loss was proposed to account for heavy mass loss years before core collapse SNe.
\cite{Soker2013} raised the possibility that the pre-explosion outburst (PEO) of the type IIn supernova 2010mc (PTF 10tel; \citealt{Ofeketal2013a})
was energized by mass accretion onto an O main-sequence stellar companion, and the ejecta had a bipolar structure.
\cite{Soker2013} conjectured that all Type IIn supernovae owe their dense circumstellar gas to binary interaction.
 \cite{Humphreysetal2012} already mentioned that SN~2011ht, as well as other types of IIn SNe, can be explained if they are
strongly non-spherical. 
It was recently proposed \citep{Franssonetal2014} that the massive ejecta \citep{Ofeketal2013b} of the PEO of SN 2010jl has a bipolar structure.
If holds, a bipolar structure strengthens the binary conjecture of \cite{Soker2013}.
The problems we find in some single-star models, further strengthen the binary model for ILOTs \citep{KashiSoker2010b}.

We thank the referee, Kris Davidson, for his helpful comments.
This research was supported by the Asher Fund for Space Research at the Technion, The US - Israel Binational Science Foundation,
and a generous grant from the president of the Technion Prof. Peretz Lavie.


\begin{thebibliography}

 \bibitem[Akashi \& Soker(2013)]{AkashiSoker2013} Akashi, M., \& Soker, N.\ 2013, \mnras, 2394

\bibitem[Arbour \& Boles(2008)]{ArbourBoles2008} Arbour, R., \& Boles, T.\ 2008, CBET, 1234, 1

\bibitem[Berger et al.(2009)]{Berger2009} Berger, E., Soderberg, A. M., Chevalier, R. A., et al. 2009, \apj, 699, 1850

\bibitem[Bond et al.(2009)]{Bond2009} Bond, H.~E., Bedin, L.~R., Bonanos, A.~Z., Humphreys, R.~M., Monard, L.~A.~G.~B.,
               Prieto, J.~L., \& Walter, F.~M.\ 2009, \apjl, 695, L154

\bibitem[Botticella et al.(2009)]{Botticella2009} Botticella, M.~T., Pastorello, A., Smartt, S.~J., et al.\ 2009, \mnras, 398, 1041

\bibitem[Boumis \& Meaburn(2013)]{BoumisMeaburn2013} Boumis, P., \& Meaburn, J.\ 2013, \mnras, 430, 3397

\bibitem[Chevalier(1976)]{Chevalier1976} Chevalier, R.~A.\ 1976,
\apj, 207, 872

\bibitem[Chevalier \& Soker(1989)]{ChevalierSoker1989} Chevalier, R.~A., \& Soker, N.\ 1989, \apj, 341, 867

\bibitem[Davidson \& Humphreys(1997)]{DavidsonHumphreys1997} Davidson, K., \& Humphreys, R.~M.\ 1997, \araa, 35, 1

\bibitem[Dessart et al.(2010)]{Dessartetal2010} {{{ Dessart, L., Livne, E.,  \& Waldman, R.\ 2010, \mnras, 405, 2113 }}}

\bibitem[Fransson et al.(2014)]{Franssonetal2014} Fransson, C., Ergon, M., Challis, P.~J., et al.\ 2014, arXiv:1312.6617

\bibitem[Humphreys et al.(2012)]{Humphreysetal2012} Humphreys, R.~M., Davidson, K., Jones, T.~J., Pogge, R. W., Grammer, S. H.,
      Prieto, J. L., \& Pritchard, T. A.\ 2012, \apj, 760, 93

\bibitem[Kashi et al.(2010)]{Kashi2010} Kashi, A., Frankowski, A., \& Soker, N.\ 2010, \apjl, 709, L11

\bibitem[Kashi \& Soker(2010a)]{KashiSoker2010a} Kashi, A., \& Soker, N.\ 2010a, \apj, 723, 602

\bibitem[Kashi \& Soker(2010b)]{KashiSoker2010b} Kashi, A., \& Soker, N.\ 2010b, (arXiv:1011.1222)

\bibitem[Kasliwal(2011)]{Kasliwal2011} Kasliwal, M.~M.\ 2011, Bulletin of the Astronomical Society of India, 39, 375

\bibitem[Kasliwal et al.(2011)]{Kasliwaletal2011} Kasliwal, M.~M., et al. 2011, \apj, 730, 134

\bibitem[Katz et al.(2012)]{Katzetal2012} Katz, B., Sapir, N., \& Waxman, E.\ 2012, \apj, 747, 147

\bibitem[Kochanek(2011)]{Kochanek2011} Kochanek, C.~S.\ 2011, \apj, 741, 37

\bibitem[Kulkarni \& Kasliwal(2009)]{KulkarniKasliwal2009} Kulkarni, S.~R., \& Kasliwal, M.~M.\ 2009, astro2010:
                    The Astronomy and Astrophysics Decadal Survey, 2010, 165

\bibitem[Mason et al.(2010)]{Mason2010} Mason, E., Diaz, M., Williams, R.~E., Preston, G., \& Bensby, T.\ 2010, \aap, 516, A108

\bibitem[Mauerhan et al.(2013)]{Mauerhanetal2013} Mauerhan, J.~C., Smith, N., Filippenko, A.~V., et al.\ 2013, \mnras, 430, 1801

\bibitem[Monard(2008)]{Monard2008} Monard, L.~A.~G.\ 2008, \iaucirc, 8946, 1

\bibitem[Mould et al.(1990)]{Mould1990} Mould, J., et al. 1990, \apjl, 353, L35

\bibitem[Ofek et al.(2008)]{Ofek2008} Ofek, E.~O., Kulkarni, S.~R., Rau, A., et al.\ 2008, \apj, 674, 447

\bibitem[Ofek et al.(2013a)]{Ofeketal2013a} Ofek, E.~O., Sullivan, M., Cenko, S.~B., et al.\ 2013a, \nat, 494, 65

\bibitem[Ofek et al.(2013b)]{Ofeketal2013b} Ofek, E.~O., Zoglauer, A., Boggs, S.~E., et al.\ 2013b, arXiv:1307.2247

\bibitem[Pastorello et al.(2010)]{Pastorello2010} Pastorello, A., et al. 2010, \mnras, 408, 181

\bibitem[Paxton et al.(2011)]{Paxton2011} Paxton, B., Bildsten, L., Dotter, A., et al.\ 2011, \apjs, 192, 3

\bibitem[Prieto et al.(2009)]{Prieto2009} Prieto, J.~L., Sellgren, K., Thompson, T.~A., \& Kochanek, C.~S.\ 2009, \apj, 705, 1425

\bibitem[Quataert \& Shiode(2012)]{QuataertShiode2012} Quataert, E., \& Shiode, J.\ 2012, \mnras, 423, L92

\bibitem[Rau et al.(2007)]{Rau2007} Rau, A., Kulkarni, S.~R., Ofek, E.~O., \& Yan, L.\ 2007, \apj, 659, 1536

\bibitem[Shiode \& Quataert(2014)]{ShiodeQuataert2014} Shiode, J.~H., \& Quataert, E.\ 2014, \apj, 780, 96

\bibitem[Smith et al.(2009)]{Smithetal2009} Smith, N., Ganeshalingam, M., Chornock, R., et al. 2009, \apjl, 697, L49

\bibitem[Smith et al.(2011)]{Smithetal2011} Smith, N., Li, W., Silverman, J.~M., Ganeshalingam, M., \& Filippenko, A.~V.\ 2011, \mnras, 415, 773

\bibitem[Soker(2013)]{Soker2013} Soker, N.\ 2013, arXiv:1302.5037

\bibitem[Soker \& Kashi(2011)]{SokerKashi2011} Soker, N., \& Kashi, A.\ 2011 (arXiv:1107.3454)

\bibitem[Soker \& Kashi(2012)]{SokerKashi2012} Soker, N., \& Kashi, A.\ 2012, \apj, 746, 100

\bibitem[Soker \& Kashi(2013)]{SokerKashi2013} Soker, N., \& Kashi, A.\ 2013, \apjl, 764, L6

\bibitem[Thompson et al.(2009)]{Thompsonetal2009} Thompson, T.~A., Prieto, J.~L., Stanek, K.~Z., Kistler, M. D., Beacom, J. F.,
         Kochanek, C. S.\ 2009, \apj, 705, 1364

\bibitem[Tsebrenko \& Soker(2013)]{TsebrenkoSoker2013} Tsebrenko, D., \& Soker, N.\ 2013, \apjl, 777, L35

\bibitem[Tylenda et al.(2013)]{Tylendaetal2013} Tylenda, R., Kaminski, T., Udalski, A., et al.\ 2013, arXiv:1304.1694

\end{thebibliography}
\end{document}